

\input phyzzx.tex
\overfullrule0pt


\def\etal{{\it et al.}}
\def\half{{\textstyle{1 \over 2}}}

\def\eighth{{\textstyle{1 \over 8}}}

\def\bold#1{\setbox0=\hbox{$#1$}%
     \kern-.025em\copy0\kern-\wd0
     \kern.05em\copy0\kern-\wd0
     \kern-.025em\raise.0433em\box0 }
\Pubnum={VAND-TH-93-11}
\date={August 1993}
\pubtype{}
\titlepage


\vskip1cm
\title{\bf The Effect of Supersymmetric Particles on
${\bold{b\rightarrow s\gamma}}$ Decay in Supergravity Models${}^*$}
\author{Marco Aurelio D\'\i az }
\vskip .1in
\centerline{Department of Physics and Astronomy}
\centerline{Vanderbilt University, Nashville, TN 37235}
\vskip .2in

\centerline{\bf Abstract}
\vskip .1in

The effect of loops involving charginos with up-type squarks, and gluinos
with down-type squarks, on the inclusive decay mode $b\rightarrow s\gamma$
is studied in the context of minimal $N=1$ Supergravity models with a
radiatively broken electroweak symmetry group. It is confirmed that the
strong constraints imposed by the CLEO upper bound $B(b\rightarrow
s\gamma)<5.4\times 10^{-4}$ on two-Higgs doublets models are much weaker
in supersymmetric theories due to partial cancelations from loops
involving charginos and up-type squarks. The dependence of the branching
ratio and the supersymmetry masses on the top quark mass is explored.

\vskip 2.cm
* Presented at the Workshop on Physics at Current Accelerators
and the Supercollider, Argonne National Laboratory, June 2-5 1993.
\vfill

\endpage

\voffset=-0.2cm

\REF\InamiL{T. Inami and C.S. Lim, {\it Prog. Theor. Phys.} {\bf 65},
297 (1981).}
\REF\TwoHDM{T.G. Rizzo, {\it Phys. Rev. D} {\bf 38}, 820 (1988); B.
Grinstein and M.B. Wise, {\it Phys. Lett. B} {\bf 201}, 274 (1988);
W.-S. Hou and R.S. Willey, {\it Phys. Lett. B} {\bf 202}, 591 (1988);
T.D. Nguyen and G.C. Joshi, {\it Phys. Rev. D} {\bf 37}, 3220 (1988);
C.Q. Geng and J.N. Ng, {\it Phys. Rev. D} {\bf 38}, 2857 (1988);
D. Ciuchini, {\it Mod. Phys. Lett. A} {\bf 4}, 1945 (1989);
B. Grinstein, R. Springer and M. Wise, {\it Nucl. Phys.} {\bf B339},
269 (1990); V. Barger, J.L. Hewett and R.J.N. Phillips, {\it Phys.
Rev. D} {\bf 41}, 3421 (1990), and Erratum.}

The decay $b\rightarrow s\gamma$ is forbiden at tree level but
induced in the Standard Model (SM) at one loop by $W$
and Goldstone bosons together with up-type quarks in the internal
lines of the loop\refmark\InamiL. The SM value of the branching
ratio of this decay is $B(b\rightarrow s\gamma)\approx
4\times 10^{-4}$ for $m_t=140$ GeV and increases with $m_t$.
In two-Higgs-doublets models, loops involving charged Higgs bosons
and up-type quarks have to be added\refmark\TwoHDM. The contribution
from the charged Higgs boson in type II models  (where one
Higgs doublet couples to the up-type quarks and the other
Higgs doublet couples to the down-type quarks)
has the same sign as the
SM contribution. In type I models (where only one Higgs doublet couples
to the fermions) the charged Higgs boson contribution does not have a
definite sign.

\REF\expbsf{E. Thorndike, CLEO Collaboration, talk given at the
1993 Meeting of the American Physical Society, Washington D.C.,
April 1993.}
\REF\hewettBBP{J.L. Hewett, {\it Phys. Rev. Lett.} {\bf 70}, 1045
(1993); V. Barger, M.S. Berger and R.J.N. Phillips, {\it Phys. Rev.
Lett.} {\bf 70}, 1368 (1993).}
\REF\misiak{M. Misiak, {\it Nucl. Phys.} {\bf B393}, 23 (1993).}
\REF\diazbtosf{M.A. D\'\i az, {\it Phys. Lett. B} {\bf 304},
278 (1993).}
\REF\ChargedH{J.F. Gunion and A. Turski, {\it Phys. Rev. D} {\bf 39},
2701 (1989);{\bf 40}, 2333 (1989); A. Brignole, J. Ellis, G. Ridolfi
and F. Zwirner, {\it Phys. Lett. B} {\bf 271}, 123 (1991); M. Drees
and M.M. Nojiri, {\it Phys. Rev. D} {\bf 45}, 2482 (1992);
A. Brignole, {\it Phys. Lett. B} {\bf 277}, 313 (1992);
P.H. Chankowski, S. Pokorski and J. Rosiek,
{\it Phys. Lett. B} {\bf 274}, 191 (1992);
M.A. D\'\i az and H.E. Haber, {\it Phys. Rev. D}
{\bf 45}, 4246 (1992).}
\REF\diaz{M.A. D\'\i az, {\it Phys. Rev. D} {\bf 48}, 2152 (1993);
M.A. D\'\i az, {\it The Fermilab Meeting DPF'92}, ed. by C. Albright
\etal, World Scientific, page 1194.}

The latest experimental upper bound on the branching fraction
for the inclusive decay mode $b\rightarrow s\gamma$, given by
$B(b\rightarrow s\gamma)<5.4\times 10^{-4}$ at 90\% c.l.
\refmark\expbsf, sets powerful constraints on the charged
Higgs boson mass in two Higgs doublets models of type II
\refmark{\hewettBBP}. Other corrections
that may be important have been calculated recently: next-to-leading
logarithmic QCD-corrections\refmark\misiak, and electroweak
corrections in the context of supersymmetry\refmark\diazbtosf
to the charged Higgs mass\refmark\ChargedH and to the charged
Higgs-fermion-fermion vertex\refmark\diaz.

\REF\BBMR{S. Bertolini, F. Borzumati, A. Masiero and G. Ridolfi,
{\it Nucl. Phys.} {\bf B353}, 591 (1991).}
\REF\susy{S. Bertolini, F. Borzumati and A. Masiero, {\it Nucl. Phys.}
{\bf B294}, 321 (1987); S. Bertolini, F.M. Borzumati, and A. Masiero,
{\it Phys. Lett. B} {\bf 192}, 437 (1987).}
\REF\Oshimo{N. Oshimo, report No. IFM 12/92, 1992, unpublished.}
\REF\Barbieri{R. Barbieri and G.F. Giudice, report No.
CERN-TH.6830/93, March 1993, unpublished.}
\REF\Nanopoulos{J.L. Lopez, D.V. Nanopoulos, and G.T. Park, report
No. CTP-TAMU-16/93, April 1993, unpublished.}
\REF\guide{J.F. Gunion, H.E. Haber, G. Kane and S. Dawson,
{\it The Higgs Hunter's Guide} (Addison-Wesley, Reading MA, 1990).}

In supersymmetry, the contributions from charginos together with
up-type squarks and from neutralinos and gluinos together with
down-type squarks, have to be included\refmark{\BBMR,\susy}.
It was stressed that in this case, it is important
the effect of loops involving
supersymmetric particles\refmark{\Oshimo-\Nanopoulos}. Here
we study this effect in the context of the radiatively
broken Minimal Supersymmetric Model\refmark\guide,
following ref. [\BBMR] and including the effects
described in refs. [\misiak,\diazbtosf].

\REF\susyrep{H.P. Nilles, {\it Phys. Rep} {\bf 110}, 1 (1984);
H.E. Haber and G.L. Kane, {\it Phys. Rep.} {\bf 117}, 75 (1985).}

Minimal $N=1$ supergravity\refmark\susyrep is characterized
by the superpotential
$$W=h_U^{ij}Q_iU_j^cH_2+h_D^{ij}Q_iD_j^cH_1+h_E^{ij}L_iE_j^cH_1
+\mu\varepsilon_{ab}H_1^aH_2^b\eqn\suppot$$
where $i,j=1,2,3$ are indeces in generation space, $\varepsilon_{ab}$
with $a,b=1,2$ is the antisymmetric tensor in two dimensions,
and $\mu$ is the Higgs mass parameter.
The different superfields transform under $SU(3)\times SU(2)\times U(1)$
according to: $Q=(3,2,1/6)$, $U^c=(\bar 3,1,-2/3)$, $D^c=(\bar 3,1,1/3)$,
$L=(1,2,-1/2)$, $E^c=(1,1,1)$, $H_1=(1,2,-1/2)$, and $H_2=(1,2,1/2)$.
The $3\times 3$ matrices $h_U, h_D$ and $h_E$ are the Yukawa couplings
given by
$$h_U={{gV^{\dagger}}\over{\sqrt{2}m_Ws_{\beta}}}\left[\matrix{m_u&0&0\cr
0&m_c&0\cr0&0&m_t\cr}\right],\qquad h_D={g\over
{\sqrt{2}m_Wc_{\beta}}}\left[\matrix{m_d&0&0\cr0&m_s&0\cr0&0&m_b\cr}
\right]\eqn\yukawaq$$
for the quarks, and for the charged leptons, $h_E$ is given by $h_D$
when the quark masses are replaced by $m_e, m_{\mu}$ ans $m_{\tau}$. The
Cabibbo-Kobayashi-Maskawa matrix is represented by $V$.
The soft supersymmetry breaking terms are
$${\cal L}_s=A_U^{ij}h_U^{ij}\tilde Q_i\tilde U^c_jH_2+A_D^{ij}
h_D^{ij}\tilde Q_i\tilde D_j^cH_1+A_E^{ij}h_E^{ij}\tilde L_i
\tilde E_j^cH_1+B\mu\varepsilon_{ab}H_1^aH_2^b+h.c.\eqn\softab$$
plus a set of scalar and gaugino mass terms, which at the unification
scale are
$${\cal L}_m=m_0^2\sum_i|S_i|^2+\big[\half M_{1/2}(\lambda_1
\lambda_1+\lambda_2\lambda_2+\lambda_3\lambda_3)+h.c.\big]
\eqn\softmass$$
where $S_i$ are all the scalars of the theory and $\lambda_i, i=1,2,3$
are the gauginos corresponding to the groups $U(1), SU(2)$ and $SU(3)$
respectively. In eq.~\softab\ all the fields are scalar components of
the respective superfields.
The mass parameters $A$ and $B$ are of ${\cal O}(m_0)$ and in some
supergravity models they satisfy the following relation at the scale
$M_X$: $A=B+m_0$
where $A$ is the common value for the $A_a^{ii}$ ($a=U,D,E$) parameters
at the unification scale: $A_U^{ij}=A_D^{ij}=A_E^{ij}=A\delta^{ij}$.

\REF\KLNPY{S. Kelley, J. Lopez, D. Nanopoulos, H. Pois and K. Yuan,
{\it Nucl. Phys.} {\bf B398}, 3 (1993).}
\REF\diazhaberii{M.A. D\'\i az and H.E. Haber,
{\it Phys. Rev. D} {\bf 46}, 3086 (1992).}

At the weak scale, the tree level Higgs potential is given by
$$\eqalign{V=&m_{1H}^2|H_1|^2+m_{2H}^2|H_2|^2-m_{12}^2(H_1H_2+h.c.)
\cr&+\eighth(g^2+g'^2)(|H_1|^2-|H_2|^2)^2+\half g^2|H_1^*H_2|^2
\cr}\eqn\Vtree$$
where $m_{iH}^2=m_i^2+|\mu|^2$ ($i=1,2$) and
$m_{12}^2=-B\mu$. The two Higgs doublets mass parameters $m_1$ and
$m_2$ satisfy $m_1=m_2=m_0$ at the unification scale $M_X$.
The three mass parameters in eq.~\Vtree\ can be replaced by the $Z$
boson mass $m_Z$, the CP-odd Higgs mass $m_A$, and the ratio between
the vacuum expectation values of the two Higgs doublets $\tan\beta
\equiv v_2/v_1$, according to the formulas
$$\eqalign{m_{1H}^2=&-\half m_Z^2c_{2\beta}+\half m_A^2(1-c_{2\beta})
\cr m_{2H}^2=&\half m_Z^2c_{2\beta}+\half m_A^2(1+c_{2\beta})\cr
m_{12}^2=&\half m_A^2s_{2\beta}\cr}\eqn\conversion$$
where $s_{2\beta}$ and $c_{2\beta}$ are sine and cosine functions of the
angle $2\beta$. The previous relations are valid at tree level. The
effects of the one-loop corrected Higgs potential may be important in
some cases\refmark\KLNPY, especially near $\tan\beta=1$ when $m_{1H}=
m_{2H}=m_{12}$ and the lightest neutral Higgs mass comes only from
radiative corrections\refmark\diazhaberii.

\FIG\susyrmi{Evolution of the different parameters of the model from
the unification scale $M_X=10^{16}$ GeV to the weak scale $m_Z$.}

\REF\rgesol{G.G. Ross and R.G. Roberts, {\it Nucl. Phys.} {\bf B377},
571 (1992).}
\REF\falck{N.K. Falck, {\it Z. Phys. C} {\bf 30}, 247 (1986).}

In Fig. \susyrmi\ it is ploted a typical solution of the renormalization
group equations (RGE) in the spirit of ref.~[\rgesol], but including the
trilinear $A$ parameters and other Higgs mass parameters as well. The
effects of the supersymmetric threshold are neglected. Also, the squark
and slepton mass parameters $M_i, i=Q,U,D,L,E$ and the $A_i, i=U,D,E$
parameters
are $3\times3$ matrices in generation space and the third diagonal
element is plotted. The RGE
used are given in ref.~[\BBMR]\ with the
exception of the $A$ parameters, whose equations are taken from
ref.~[\falck]. The set of independent parameters is chosen to be $m_t,
m_A$ and $\tan\beta$ at the weak scale, $M_{1/2}$ at the unification
scale, and the sign of $\mu$ as a discrete parameter.
Several features can be seen from Fig.~\susyrmi. The sleptons are lighter
than the squarks, the reason being that there is no contribution from
the strong coupling constant to their RGE. Besides, $M_L$ is larger than
$M_E$ at the weak scale because the gauge coupling constant $g$ is larger
than $g'$. In the squark mass parameters, the effect of the gauge
couplings is counteracted by the Yukawa couplings, especially from the
top quark. $M_D$ is the largest because it does not receive an opposite
effect from the top quark Yukawa; on the other hand, $M_U$ is the smaller
of the three squark mass parameters because in its RGE the top quark
Yukawa coupling
has a larger coefficient. The evolution of the mass parameters in the
gaugino sector is trivial since it is governed by the respective gauge
coupling. In the Higgs sector, since the squared masses may became
negative, $-\sqrt{-m^2_i}$ is plotted when $m^2_i<0$, where $i=
1H, 2H, 12$. The evolution of $m_1$ is dominated by the gauge coupling
constants, but since there is no contribution from the strong coupling,
its dependence on the scale is weak. On the contrary, the evolution of
$m_2$ is dominated by the top Yukawa coupling, producing a large splitting
$m_1^2-m_2^2$ at the weak scale. The value of this splitting is a
necesary ingredient to produce the correct electroweak symmetry breaking.

The QCD uncorrected amplitude for the decays $b\rightarrow s\gamma$
and $b\rightarrow sg$ are
$$A^{\gamma,g}(m_W)=A_{SM}^{\gamma,g}+(f^+f^-)
A_{H^{\pm}}^{1\gamma,g}+(f^-)^2\cot^2\beta A_{H^{\pm}}^{2\gamma,g}+
A_{\tilde\chi^{\pm}}^{\gamma,g}+A_{\tilde g}^{\gamma,g}
\eqn\Atotfg$$
where the form factors $f^{\pm}$ come from the renormalization
of the charged Higgs boson coupling to a pair of fermions\refmark\diaz.
The different amplitudes $A$ can be found in ref.~[\BBMR].
If we now run the scale from $m_W$ to $m_b$ and introduce the QCD
corrections, we get
$$A^{\gamma}(m_b)=\eta^{-{\textstyle{16\over 23}}}\Bigg[
A^{\gamma}(m_W)+{8\over 3}A^g(m_W)(\eta^{\textstyle{2\over 23}}-1)
\Bigg]+CA_0^{\gamma}\eqn\AcorrQCD$$
where $\eta=\alpha_s(m_b)/\alpha_s(m_W)\approx 1.83$ and $A_0^{\gamma}$
is given by
$$A_0^{\gamma}={{\alpha_W\sqrt{\alpha}}\over{2\sqrt{\pi}}}{{V_{ts}^*
V_{tb}}\over{m_W^2}}\eqn\azero$$
with $C=0.177$, $\alpha_W=g^2/4\pi$, and $\alpha=e^2/4\pi$. This last
term proportional to
$C$ comes from mixing of four quark operators\refmark\Barbieri.

\FIG\bsfiii{Dependence on the top quark mass of:
(a) Branching ratio of the inclusive decay $B(b\rightarrow
s\gamma)$ for the SM and for the SUSY-GUT model.
(b) Relative size of the chargino, charged Higgs and gluino
contributions to the $b\rightarrow s\gamma$ amplitud with respect to the
SM amplitud. The QCD corrections are not implemented. Note that the sign
of the chargino and gluino contributions is changed, and that the
gluino contribution is amplified by a factor of 10.
(c) Masses of the lightest and the heviest up-type squark
(solid) and charginos (dashes), the charged Higgs
(dotdash) and the gluino (dots).
(d) Masses of the lightest and the heviest down-type squark
(solid), charged slepton (dashes) and neutralino (dotdash), and the
lightest of the sneutrinos (dots).
(e) Common scalar mass parameter $m_0$ at the unification
scale and the Higgs mass parameters $B$ and $\mu$ at the weak scale.}

In fig.~\bsfiii\ the top quark mass is taken as a variable, keeping all
the other parameters of fig.~\susyrmi\ unchanged. The dependence on the
top quark mass of the branching ratio $B(b\rightarrow s\gamma)$ can be
seen in fig.~\bsfiii(a). In the SM this branching ratio grows with the top
quark mass and remains below the CLEO bound in the hole range of $m_t$.
In the SUSY--GUT model, for the parameters considered here, the branching
ratio decreases with the top quark mass exept for very large values of
$m_t$. This effect is due to a faster growing (and with opposite sign)
chargino contribution compared to the charged Higgs contribution
[see fig.~\bsfiii(b)]. This can be seen also in ref.~[\Barbieri]
taking into account that in the SUSY--GUT model, a change in $m_t$ implies
a change in $m_0$ [see fig.~\bsfiii(e)], in opposition to the
treatment in
ref.~[\Barbieri] where both parameters are independent. If the top quark
mass increases, the splitting $m_1^2-m_2^2$ at the weak scale increases
also, and to keep it constant $m_0$ must decrease. This in turn will
decrease the absolute values of $m_1^2$ and $m_2^2$, so $\mu$ will grow
in order to keep $m_{1H}^2$ and $m_{2H}^2$ unchanged. Finally,
in figs.~\bsfiii(c) and~\bsfiii(d) are displayed the masses
of the different
particles as a function of the top quark mass. As a final comment, it is
clear that the decay $b\rightarrow s\gamma$ does not strongly constrain
supersymmetry, and in the case of minimal $N=1$ supergravity models with
radiatively broken electroweak symmetry, the branching ratio $B(b
\rightarrow s\gamma)$ lies below the CLEO bound and even below the
SM value for reasonable values of the free parameters.

\REF\GnO{Y. Okada, report No. KEK-TH-365, July 1993, unpublished;
R Garisto and J.N. Ng, report No. TRI-PP-93-66, July 1993, unpublished.}

\vskip .5cm
\noindent{\bf Note added:}
When this work was completed, we received two preprints\refmark\GnO
that calculate the effect on $B(b\rightarrow s\gamma)$ due to loops
involving charginos and up-type squarks.
\vskip .5cm

\vskip .5cm
\centerline{\bf ACKNOWLEDGMENTS}
\vskip .5cm

Discussions with Howard Baer, Joseph Lykken, Xerxes Tata, and Thomas
Weiler are gratefully acknowledged.
I am thankful to the Particle Theory Group at Fermilab, where part of
this work was completed.
This work was supported by the U.S. Department of Energy, grant No.
DE-FG05-8SER40226.

\refout
\figout
\end